\documentclass[12pt, reqno]{amsart}
\usepackage{amsmath, amsthm, amscd, amsfonts, amssymb, graphicx, color}
\usepackage[bookmarksnumbered, colorlinks, plainpages,
    citecolor=blue ]{hyperref}
\usepackage{textcase}
\usepackage{tikz}
\usetikzlibrary{positioning, shapes.geometric, arrows.meta, fit}
\usepackage{caption}
\usepackage{enumitem}
\usepackage{float}

\usepackage{booktabs}
\usepackage[numbers]{natbib}

\usepackage{listings}
\usepackage{xcolor}
\theoremstyle{definition}

\theoremstyle{remark}

\numberwithin{equation}{section}

\makeatletter
\def\@secnumfont{\bfseries}
\renewcommand{\section}{\@startsection{section}{1}{\z@}%
  {-3.5ex \@plus -1ex \@minus -.2ex}%
  {2.3ex \@plus.2ex}%
  {\normalfont\large\bfseries}}
\renewcommand{\subsection}{\@startsection{subsection}{2}{\z@}%
  {-3.25ex\@plus -1ex \@minus -.2ex}%
  {1.5ex \@plus .2ex}%
  {\normalfont\bfseries}}
\renewcommand{\subsubsection}{\@startsection{subsubsection}{3}{\z@}%
  {-3.25ex\@plus -1ex \@minus -.2ex}%
  {1.5ex \@plus .2ex}%
  {\normalfont\itshape}}
\makeatother

\newcommand{\titlecite}[1]{%
  {\NoHyper\citeauthor{#1} (\citeyear{#1})~\endNoHyper}\citep{#1}%
}

\begin{document}

\pagestyle{plain}
\setcounter{page}{1}

\title{HOT Protocol}

\author{Peter Volnov, Georgii Kuksa \MakeLowercase {and} Andrey Zhevlakov\\
{\normalfont\itshape\NoCaseChange{HOT Labs}}}

\begin{abstract}

	HOT Protocol provides the infrastructure that allows smart contracts
	to securely own and manage private keys.
	The Multi-Party Computation (MPC) Network manages signing keys.
	By running an MPC node inside the Trusted Execution Environment (TEE),
	the protocol achieves stronger security guarantees while
	lowering economic requirements for participants.
	The NEAR Protocol provides a decentralized and efficient state layer.
	Key management can be integrated with any smart contract
	across Stellar, TON, Solana, and EVM-compatible networks.

\end{abstract}

\maketitle

\tableofcontents


\section{Introduction}\label{sec:introduction}

Smart contracts are restricted to modifying and verifying
state within their native blockchain,
resulting in a fragmented Web3 ecosystem.
Enabling smart contracts to cryptographically
sign messages under their own identity would allow autonomous
control over actions extending beyond the boundaries of a single
blockchain domain.

This design introduces two challenges:
\begin{enumerate}
	\item Message signing must be externally triggered,
	      since smart contracts act only upon incoming transactions

	\item A smart contract cannot securely store private keys,
	      as its state is publicly accessible
\end{enumerate}

To resolve the first issue, a pull-based approach allows any participant to request
a signature from a contract-associated key,
provided that the contract explicitly authorizes the message.

The second challenge can be solved with a Multi-Party
Computation (MPC) network.
It securely creates and uses a secret key for signing messages,
and any operation with that key requires approval from a threshold
number of participants.

Sections 2–4 describe implementation details and their interconnection,
and provide developers with instructions on enabling key management,
as well as an overview of the system’s trust and threat model.

\section{System Design}\label{sec:system-design}

We first present the communication model and the primary actors in the signature generation
process, then describe each actor’s behavior and implementation.

\begin{tikzpicture}[
	font=\large,
	>=Stealth,
	box/.style={
			draw, rounded corners=6pt, line width=0.8pt,
			minimum width=4.5cm, minimum height=1.8cm,
			align=center
		},
	arrow/.style={line width=0.9pt, -{Stealth[length=3mm]}},
	lbl/.style={font=\large, inner sep=1pt}
	]

	\node[box] (third) at (0,0) {Third-party};
	\node[box] (mpc) at (6,0) {MPC\\Network};
	\node[box] (key) at (12,0) {Key-Owner\\Contract};

	\draw[arrow, bend left=25]
	(third.north east) to node[lbl, above, pos=0.5]{1} (mpc.north west);
	\draw[arrow]
	(mpc.east) -- node[lbl, above, pos=0.5]{2} (key.west);
	\draw[arrow, bend left=25]
	(mpc.south west) to node[lbl, below, pos=0.5]{3} (third.south east);

\end{tikzpicture}


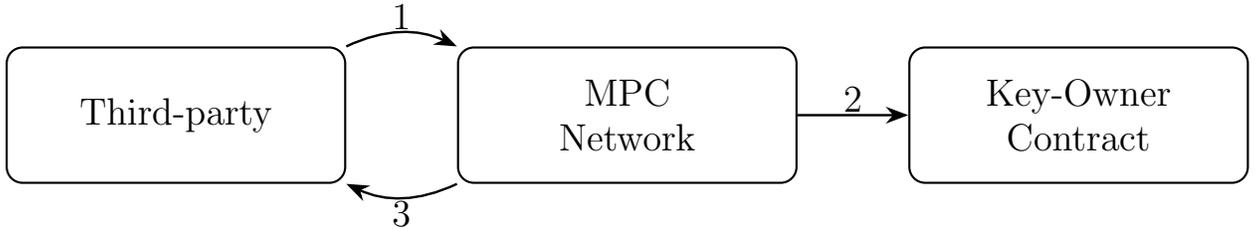
\captionof{figure}{\parbox{0.8\textwidth}{
		High-level flow for obtaining a signature from a smart contract.
	}}

\label{fig:system-flow}

\begin{enumerate}
	\item A third party triggers the pull mechanism,
	      attempting to obtain a signature for a message from
	      a private key linked to a specific smart contract.
	      A smart contract may be linked to the multiple private keys.

	\item The MPC network checks whether the smart contract authorizes the
	      message to be signed.
	      This check is read-only and leaves the contract state unchanged.

	\item The MPC network securely signs the message and sends
	      the result back to the third party.

\end{enumerate}

\subsection{Key-Owner Contract}\label{subsec:key-owner-contract}

This section explains the steps a developer needs to follow
to enable key management for their contract.

\begin{figure}[H]
	\begin{tikzpicture}[
		font=\large, >=Stealth,
		box/.style={draw, rounded corners=6pt, line width=0.8pt,
				minimum width=4.2cm, minimum height=1.8cm, align=center},
		arrow/.style={line width=0.9pt, -{Stealth[length=3mm]}},
		lbl/.style={font=\large, inner sep=1pt}
		]

		\node[box] (keyowner) at (0,1.5) {Key-Owner\\Contract};
		\node[box] (mpc)      at (0,-1.5) {MPC\\Network};

		\node[box] (auth)   at (6,1.5) {chain\_id,\\contract\_address};
		\node[box] (pubkey) at (6,-1.5) {\texttt{key\_id}};

		\node[box] (registry) at (12,0) {Key\\Registry};

		\coordinate (merge) at (9,0);

		\draw[arrow] (keyowner.east) -- node[lbl, above, pos=0.5]{1} (auth.west);
		\draw[arrow] (mpc.east)      -- node[lbl, above, pos=0.5]{2} (pubkey.west);

		\draw[line width=0.9pt] (auth.east)   to[out=-15,in=120,looseness=1.1] (merge);
		\draw[line width=0.9pt] (pubkey.east) to[out=15,in=-120,looseness=1.1] (merge);

		\draw[arrow] (merge) -- node[lbl, above, pos=0.55]{3} (registry.west);

	\end{tikzpicture}\label{fig:figure2}
\end{figure}

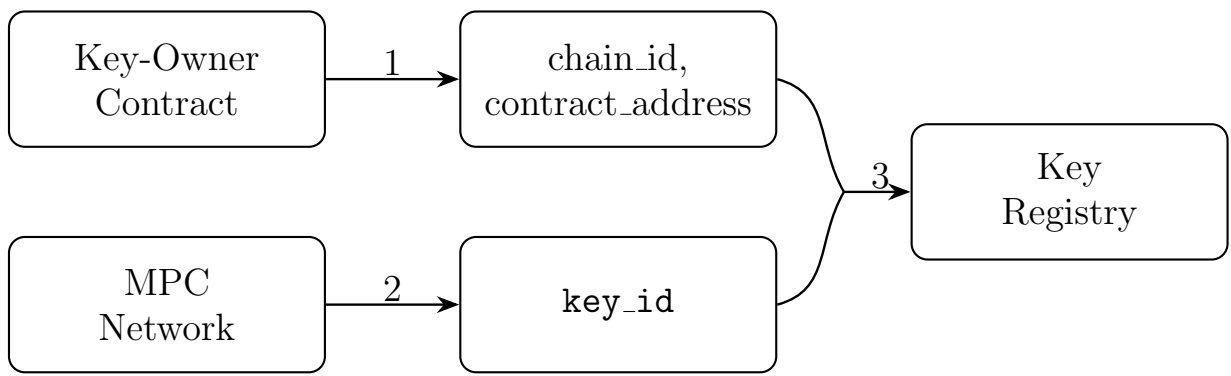
\captionof{figure}{\parbox{0.8\textwidth}{
		Key Management Setup
	}}

\begin{enumerate}
	\item
	      Key management requires the contract to define the following
	      read-only method:

	      \begin{figure}[H]
		      \begin{lstlisting}[language=C, caption={Authorization function signature},label={lst:lstlisting1}, escapeinside={(*@}{@*)}]
fn hot_verify(
	message: String,
	key_id: bytes32,
	metadata: bytes,
) -> bool;
				\end{lstlisting}\label{fig:figure}
	      \end{figure}
	      Input arguments:
	      \begin{itemize}
		      \item
		            \texttt{message} — the message being signed

		      \item
		            \texttt{key\_id} — an identifier used to distinguish
		            between multiple keys owned by the contract.
		            One contract can hold multiple keys,
		            removing the need to redeploy logic per user.

		      \item
		            \texttt{metadata} — auxiliary data used in the authorization process. \\
		            For example, if \texttt{message} is a transaction hash, the transaction pre-image is passed as \texttt{metadata} for invariant verification.

	      \end{itemize}
	      Return value: A boolean, indicating whether the message
	      is authorized for signing.

	      Each contract is uniquely defined by the pair
	      \texttt{(chain\_id, contract\_address)}.

	\item
	      A key is reserved via the MPC Network
	      (\ref{subsubsec:kdf}), returning a \texttt{key\_id}.

	\item
	      The Key Registry contract\label{par:key-registry} stores
	      the mapping
	      between \texttt{key\_id} and its authorization contract.

\end{enumerate}

After setup, the MPC network performs an authorization check for the given
\texttt{key\_id} before proceeding with signature generation, by:
\begin{enumerate}
	\item fetching its authorization contract from the Key Registry;
	\item calling \texttt{hot\_verify} with the target message.
\end{enumerate}

The Key-Owner contract should make the \texttt{key\_id} publicly
accessible (e.g., through a view method), enabling third parties to confirm
that key control is enforced by the \texttt{hot\_verify} method.

\par\medskip

Practical example of the mentioned process can be found in Appendix \ref{sec:passkey-bitcoin-wallet}.

\subsection{MPC Network}\label{subsec:mpc-network}

Once the contract layer defines how authorization is verified,
the MPC network ensures that key operations follow these authorizations.

\subsubsection{Design Principles}
The MPC network is designed around three key principles
that guide the proper use of the key management system.

\begin{itemize}
	\item Decentralization – no single entity controls
	      the private key or computation;
	      all nodes collaborate in a trust-minimized manner.
	\item Security – resistant to adversarial behavior,
	      preserving computational correctness and data confidentiality.
	\item Scalability – designed to operate efficiently as the number
	      of participating nodes or supported chains increases.
\end{itemize}
This leads to the idea that the private key used for message
signing is divided into cryptographic shares and distributed
among the participants of the MPC network.

\subsubsection{Security Guarantees}

The design results in the following properties,
contributing to the overall correctness of key management:

\begin{itemize}
	\item Threshold cooperation: a minimum number of nodes must
	      collaborate to successfully generate a valid signature.
	\item Key secrecy: no single node ever possesses the complete
	      private key; each holds only a share.
	      An adversary would need to compromise at least the threshold
	      number of nodes to reconstruct or misuse the key.
	\item Signature indistinguishability: MPC-generated signatures
	      are identical to those from the master private key,
	      unlike traditional multisignature outputs.

\end{itemize}

\subsubsection{Interface}
For the system to function properly with contract-based key management
and maintain the stated design properties,
the MPC network should be capable of performing the following operations:

\begin{itemize}
	\item Distributed Key Generation (DKG): securely generates
	      a private key and distributes its shares across participating nodes.
	      Implementation follows the protocol by \titlecite{gennaro-dkg}

	\item Key Resharing – enables the addition or removal of participants,
	      or modification of the threshold value.
	      The key remains the same, but all shares are refreshed
	      to reflect the new configuration.
	      The implemented protocol is a specific instance of the general
	      Distributed Key Generation (DKG) scheme.

	\item Message Signing — cooperatively generates a valid signature
	      when a threshold of nodes participates, without ever
	      reconstructing the private key.
	      The implementation depends on the chosen signature scheme.
	      In this work, we focus on the two most widely adopted ones:

	      \begin{itemize}
		      \item ECDSA, used in Bitcoin, EVM-compatible chains, and Tron.
		            The implementation the protocol by \titlecite{fast-ecdsa}

		      \item EdDSA, used in Solana, NEAR, Stellar, and TON\@.
		            The implementation follows the protocol by \titlecite{frost}
	      \end{itemize}
\end{itemize}

\subsubsection{Diagram: Initialization and Resharing flow}

Before enabling key management,
the MPC network needs to be initialized and configured to support participant updates.

\begin{tikzpicture}[
		box/.style={rectangle, draw, rounded corners, minimum width=2.5cm, minimum height=1.5cm, align=center, thick},
		bigbox/.style={rectangle, draw, rounded corners, thick, inner sep=15pt},
		arrow/.style={-Stealth, thick}
	]

	\node[box, minimum width=1.8cm, minimum height=1.2cm] (crypto) at (-4.5,0.0) {Crypto\\Module};
	\node[box, minimum width=2cm, minimum height=1.2cm] (blockchain) at (-2.5,0.0) {Blockchain\\Client};

	\node[bigbox, fit=(crypto)(blockchain)] (node1) at (-3.5,0) {};
	\node[anchor=north west] at (node1.north west) {\textbf{Node \#1}};

	\node[box, minimum width=1.5cm, minimum height=0.8cm] (dots) at (-3.5,-2.0) {...};
	\node[box, minimum width=1.8cm, minimum height=0.8cm] (nodeN) at (-3.5,-3.2) {Node \#N};

	\node[bigbox, fit={(node1) (dots) (nodeN)}, inner sep=20pt, label={[anchor=north west]north west:\textbf{MPC Network}}] (network) {};

	\node[box, minimum width=2.2cm, minimum height=1.3cm] (controller) at (3,0.0) {MPC\\Controller\\Contract};

	\node[anchor=west] (genesis) at (5.5,0.0) {Genesis Deploy};

	\draw[arrow] (blockchain.east) -- ++(1.2,0) node[above, midway] {2} -- (controller.west);
	\draw[arrow] (genesis.west) -- node[above] {1} (controller.east);
	\draw[arrow] (crypto.west) -- ++(-0.2,0) |- node[pos=0.1, left] {3} (nodeN.west);

\end{tikzpicture}


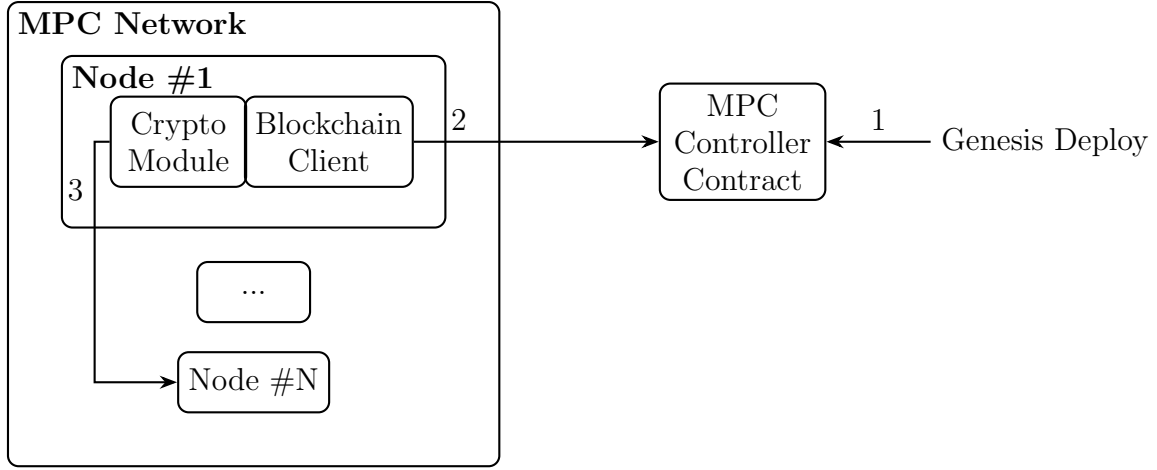
\captionof{figure}{\parbox{0.8\textwidth}{
		MPC Initialization flow
	}}

\begin{enumerate}
	\item The MPC Controller Contract is deployed
	      in the genesis block to serve two main purposes:
	      \begin{itemize}
		      \item State management: stores operational data of
		            the MPC network, including participant details like
		            IP addresses, encryption public keys, and identifiers
		      \item Network coordination: handles configuration changes
		            through on-chain voting and triggers key resharing when
		            thresholds or participant sets are updated.
	      \end{itemize}
	      The contract only needs to be cost-efficient for state storage,
	      making NEAR Protocol an appropriate choice.
	      The same applies to the
	      Key Registry contract (\ref{par:key-registry}).
	\item Nodes get their operational data from the Controller Contract.
	      To keep the diagram simple, only one arrow
	      is shown — in practice, every node fetches the data
	\item Nodes run the DKG process to generate and receive their
	      private key shares.
	      For brevity reasons a single arrow is shown.
	      In practice there is an arrow for each pair of participants.
\end{enumerate}

\subsubsection{Key Derivation Function}\label{subsubsec:kdf}

As noted in \ref{subsec:key-owner-contract}, a set of keys should be
associated with a smart contract, which can be achieved
by performing multiple DKG invocations.

To ensure scalability, the network avoids repeating DKG
and Key Resharing for every new configuration.
Instead, a single root key is established through DKG,
and all subsequent operational keys are obtained through
deterministic derivation.

Derived keys are referenced via a tweak, defined as the \texttt{key\_id}.

\par\medskip

This derivation model follows a mechanism similar to
idea by \titlecite{bip32}, where a child public key can be
derived from a parent public key using a known tweak,
without access to the private shares.

\subsubsection{Message Signing Authorization}

To produce a valid signature,
the smart contract first authorizes the message for a particular \texttt{key\_id}.

Each MPC node validates this authorization before engaging in the signing protocol by:
\begin{enumerate}[label=\alph*)]
	\item Retrieving the contract address associated with the specified \texttt{key\_id} from the Key Registry.
	\item Executing the authorization method of the key-owner contract.
\end{enumerate}
The process for (a) is described in \ref{subsec:key-owner-contract}.

For (b), each MPC node must operate a light node for every
supported network, or a full node where a light node
implementation is not available.

\subsubsection{Trusted Execution Environment}

While the system’s security primarily relies on cryptographic design,
it can be further reinforced through hardware-based mechanisms.

A Trusted Execution Environment (TEE) is a hardware-enforced
isolated execution context that offers the following guarantees:

\begin{itemize}
	\item Confidentiality: The TEE guarantees that both code and data
	      executed within an enclave remain isolated and unreadable to any
	      external software layer, preserving the secrecy of
	      sensitive computations
	\item Integrity: The TEE ensures that enclave code executes
	      as deployed and that neither its logic nor runtime state can
	      be altered by external software.
	\item Attestation: The TEE can produce cryptographic proof
	      of its software identity and state, allowing remote parties
	      to verify that it runs trusted code.
\end{itemize}

The MPC node is executed within a trusted environment,
providing isolation for high-risk operations:

\begin{itemize}
	\item Computations with private key shares
	\item Verifying contract authorization method
\end{itemize}

To maintain correctness guarantees,
each MPC node periodically provides an attestation
to the Controller Contract, which performs verification
of the reported state.

In practical terms, following technologies are being used
to provide hardware-assisted isolation at the VM layer:
\begin{itemize}
	\item TDX on Intel platforms (\titlecite{intel-tdx})
	\item SEV on AMD platforms (\titlecite{amd-sev})
\end{itemize}

Utilizing the integrity property, node logs can be treated as
a source of truth, providing a reliable basis for
the governance model, as will be described
in section \ref{sec:economic-and-governance-model}.

\subsubsection{Access Control Layer: Gatekeeper Network}

To prevent overload and malicious use of the MPC network,
a dedicated intermediary layer handles signature requests from third parties.
Consequently, only the Gatekeeper network is permitted to access
the MPC network directly.

Gatekeepers serve two primary functions:
\begin{itemize}
	\item Perform load balancing and filtering of incoming
	      requests to guarantee correctness and validity
	\item Ensures clear separation between the economic domains of MPC nodes and end users.
	      Gatekeepers never handle or store private key material.
	      Developers may customize and self-host Gatekeepers for their specific workflows,
	      even without TEE support.
	      This separation, along with the incentive model for this role,
	      is described in section \ref{subsubsec:gatekeeper}.
\end{itemize}

Offloading these two tasks to the Gatekeeper
reduces the attack surface of the MPC nodes.

\subsection{Full Signature generation flow}\label{subsec:diagram:-signature-generation-flow}

With the contract interface and MPC mechanics defined, the full signature flow is as follows:

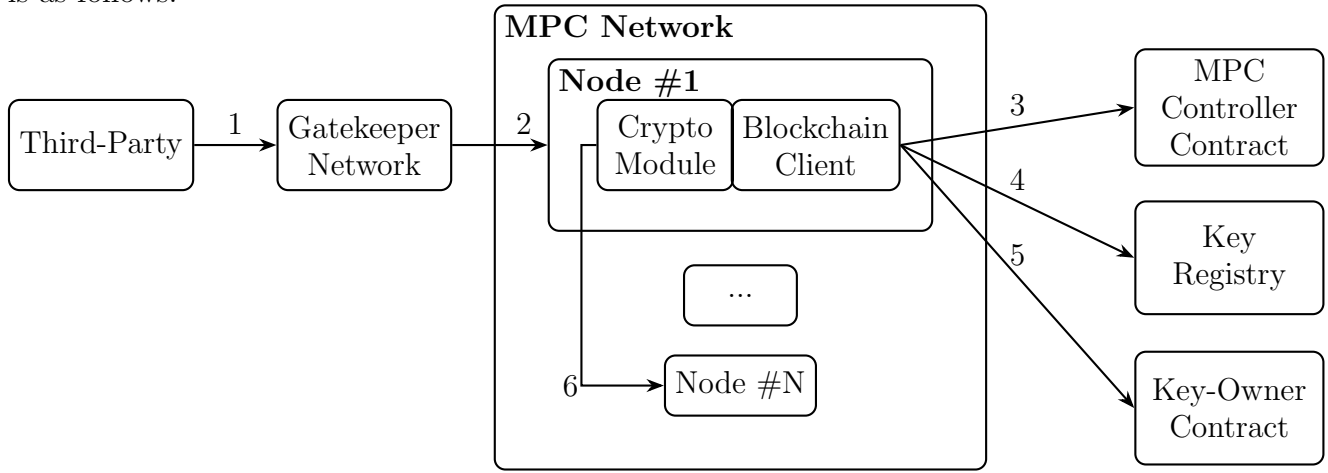
\begin{tikzpicture}[
		box/.style={rectangle, draw, rounded corners, minimum width=2.5cm, minimum height=1.5cm, align=center, thick},
		bigbox/.style={rectangle, draw, rounded corners, thick, inner sep=15pt},
		arrow/.style={-Stealth, thick}
	]

	\node[box, minimum width=1.8cm, minimum height=1.2cm] (gatekeeper) at (-8.5,0.0) {Gatekeeper\\Network};
	\node[box, minimum width=1.8cm, minimum height=1.2cm] (third-party) at (-12.0,0.0) {Third-Party};
	\node[box, minimum width=1.8cm, minimum height=1.2cm] (crypto) at (-4.5,0.0) {Crypto\\Module};
	\node[box, minimum width=2cm, minimum height=1.2cm] (blockchain) at (-2.5,0.0) {Blockchain\\Client};

	\node[bigbox, fit=(crypto)(blockchain)] (node1) at (-3.5,0) {};
	\node[anchor=north west] at (node1.north west) {\textbf{Node \#1}};

	\node[box, minimum width=1.5cm, minimum height=0.8cm] (dots) at (-3.5,-2.0) {...};
	\node[box, minimum width=1.8cm, minimum height=0.8cm] (nodeN) at (-3.5,-3.2) {Node \#N};

	\node[bigbox, fit={(node1) (dots) (nodeN)}, inner sep=20pt, label={[anchor=north west]north west:\textbf{MPC Network}}] (network) {};

	\node[box] (controller) at (3,0.5) {MPC\\Controller\\Contract};
	\node[box] (registry) at (3,-1.5) {Key\\Registry};
	\node[box] (key-owner) at (3,-3.5) {Key-Owner\\Contract};

	\draw[arrow] (third-party.east) -- node[above, midway] {1} (gatekeeper.west);
	\draw[arrow] (gatekeeper.east) -- node[above, pos=0.75] {2} (node1.west);
	\draw[arrow] (blockchain.east) -- node[above, midway] {3} (controller.west);
	\draw[arrow] (blockchain.east) -- node[above, midway] {4}  (registry.west);
	\draw[arrow] (blockchain.east) -- node[above, midway] {5}  (key-owner.west);
	\draw[arrow] (crypto.west) -- ++(-0.2,0) |- node[pos=0.55, left] {6} (nodeN.west);

\end{tikzpicture}

\captionof{figure}{\parbox{0.8\textwidth}{
		Signature generation flow
	}}

\begin{enumerate}
	\item A third party triggers the pull-based signing
	      flow via the Gatekeeper Network, requesting a signature
	      for a message tied to the public key of a given smart
	      contract and submitting the necessary inputs to its
	      authorization method(s).

	\item The Gatekeeper manages interactions with the third
	      party and chooses which threshold nodes will take part in the signing.

	\item MPC nodes retrieve protocol-specific details from one another,
	      such as IP addresses.
	      In reality, this happens only once during initialization,
	      with updates performed on each MPC configuration change.
	      For simplicity, the diagram here and in the following steps
	      shows a single arrow.

	\item MPC nodes obtain the authorization methods corresponding
	      to the public key involved in the request.

	\item MPC nodes call the read-only authorization method
	      to check whether the message is allowed to be signed.

	\item After successful authorization, the MPC nodes collaboratively generate
	      the cryptographic signature.
\end{enumerate}

While the design ensures cryptographic correctness and operational scalability,
maintaining long-term integrity requires on-chain governance and economic incentives,
discussed in the following section.

\section{Governance Model}\label{sec:economic-and-governance-model}

The protocol combines cryptographic and hardware security with economic
and governance mechanisms to guarantee accountability,
liveness, and deterministic network costs.
At the core of the architecture is the \texttt{gTOKEN},
a native asset that underpins staking, incentive alignment,
and governance execution across all protocol layers.

\subsection{Governance token}\label{subsec:governance-token}

\texttt{gTOKEN} is a fungible token on NEAR Protocol (\titlecite{nep141}).

Used to:
\begin{itemize}
	\item rent MPC capacity via Gatekeepers

	\item lock collateral for nodes and Gatekeepers

	\item distribute rewards for honest behavior.
	      Each NEAR Protocol epoch mints new \texttt{gTOKEN}s,
	      distributed according to participants’ contributions.

\end{itemize}

\subsection{Roles}\label{subsec:roles}

The following actor classes together sustain the protocol’s operational and economic integrity:
\begin{enumerate}
	\item MPC Nodes
	\item Decentralized Autonomous Organization (DAO)
	\item Gatekeepers
	\item Fishermen
	\item Smart Contract Users
\end{enumerate}

\subsubsection{DAO}
The DAO forms the top-level governance layer ensuring protocol integrity and parameter control.

DAO members are independent organizations with a material \texttt{gTOKEN} stake.
Joining requires member approval and a signed pledge to the DAO manifesto.
Members of DAO must:
\begin{itemize}
	\item Veto any decision that endangers security or stability of the MPC network
	\item Approve new DAO members that meet the requirements
	\item Approve new MPC network parameters change
	\item Approve new Gatekeepers
	\item Process reports against Gatekeepers and MPC nodes; verify TEE-signed logs; execute slashing when violations are proven
	\item Define and adjust economic parameters — including inflation rate,
	      reward schedule for MPC nodes, and minimum staking requirements for Gatekeepers
	      and node operators
\end{itemize}

\subsubsection{Gatekeeper}\label{subsubsec:gatekeeper}
Gatekeepers lease a portion of the protocol’s capacity,
defined by rate limits, and allocate that capacity to their users.
They are free to determine their own business model: charging usage fees,
accepting stablecoins,
or provide it for free for some users depending on the other factors.

Each Gatekeeper co-signs a user’s request \texttt{(gatekeeper\_id, request, deadline)}
and relays it to the MPC Network.
These signed receipts are publicly available in MPC logs and
can serve as evidence for slashing a Gatekeeper’s stake in cases of
malicious behavior or rate-limit violations.

\subsubsection{Fishermen}

Any third party can monitor public TEE-signed logs and open
disputes in the controller contract.

Dispute resolution relies on TEE-signed MPC node logs, which include:
\begin{itemize}
	\item timestamps;
	\item signing requests submitted by Gatekeepers;
	\item validation outcomes;
	\item protocol round statuses;
	\item node unavailability reports;
	\item protocol-level errors.
\end{itemize}

Submitting a dispute incurs a \texttt{gTOKEN} fee to prevent spam.
Upon successful slashing, the Fisherman is rewarded with a share of the penalized stake.

\subsubsection{MPC Node provider}
Core hardware provider. Supports the continuous and stable operation of its MPC Node and gets rewarded for this from inflation.

\begin{itemize}
	\item Stake \texttt{gTOKEN} to join the network
	\item earn \texttt{gTOKEN} after each epoch
	\item \texttt{gTOKEN} can be slashed for unstable operation
\end{itemize}

\begin{table}
	\centering
	\begin{tabular}{llll}
		\toprule
		\textbf{Role} & \textbf{Stake gTOKEN} & \textbf{Earn gTOKEN} & \textbf{Contribution} \\
		\midrule
		DAO Member

		              & YES

		              & YES

		              & Governance

		\\
		MPC Node

		              & YES

		              & YES

		              & Hardware

		\\
		Gatekeeper

		              & YES

		              & NO

		              & Distribution

		\\
		User

		              & NO

		              & NO

		              & Smart contract user

		\\
		Fishermen

		              & NO

		              & YES

		              & Monitoring

		\\
		\bottomrule
	\end{tabular}
	\caption{}
	\label{tab:}
\end{table}

With the governance framework in place, we next examine the security guarantees
and potential attack surfaces within this operational model.

\section{Security Model}\label{sec:security-model}

This section outlines the scope of attacks.
The system assumes an honest majority among MPC nodes,
verified TEE integrity, and correct operation of the on-chain governance mechanisms.
Within this model, potential adversaries are limited to those attempting
to disrupt availability or compromise confidentiality.

\subsection{Denial of Service}\label{subsec:denial-of-service}

\subsubsection{Uncooperative node}

Threat: a node stalls or refuses rounds to cause signing delays or failures.

Mitigations:
\begin{itemize}
	\item On-chain: vote to exclude the node and run key resharing;
	      slashing on proven misbehavior

	\item Operational: Gatekeeper selects any threshold
	      of responsive nodes; can blacklist slow nodes;
	      group-testing (GBS-style) selection

	\item Operational: health checks maintain an active set;
	      rolling updates avoid synchronized downtime
\end{itemize}

\subsubsection{Number of active nodes $\leq \texttt{threshold}-1$}

Mitigation:
\begin{itemize}
	\item Operational: Increase n and diversify operators and hosting across
	      independent clusters
\end{itemize}

\subsubsection{Gatekeeper quota abuse}
Threat: A malicious or compromised Gatekeeper may abuse its quota to flood the MPC network
reducing throughput and availability.

Mitigation:
\begin{itemize}
	\item Operational: Enforce per-node request rate limits and auto-eject policies;
	\item On-chain: Allow MPC nodes to collectively vote to remove the offending Gatekeeper,
	      applying slashing penalties verified via TEE-signed logs.
\end{itemize}

\subsubsection{Gatekeeper downtime}
If a Gatekeeper fails, users switch to another Gatekeeper;
no unique access is held by any single Gatekeeper.

\subsection{Confidentiality}\label{subsec:confidentiality}

\subsubsection{Mimicry of TEE}
Threat: a node pretends to run attested code.

Mitigation:
\begin{itemize}
	\item On-chain: Time-bound remote attestation, periodic re-attestation,
	      code-identity checks pinned in Coordinator;
	      nodes failing re-attestation are disabled before protocol rounds.
\end{itemize}

\subsubsection{Key leakage via RAM/Persistence}
Threat: a coalition reconstructs user keys by reading shares
from memory or disk.

Preconditions:
\begin{itemize}
	\item coalition size $\geq \texttt{threshold}$
	\item bypass TEE memory and persistence protections
\end{itemize}

Mitigation:
\begin{itemize}
	\item Operational: memory isolation via TEE enclaves
	\item On-chain: periodic re-attestation
\end{itemize}

\subsubsection{Controller or Key Registry Contract hijacking}
Mitigations:
\begin{itemize}
	\item On-chain: Both contracts operate without access keys,
	      preventing unilateral modification of authorization mappings
	      or attestation logic.
\end{itemize}

Together, these mitigations demonstrate that security properties
established in Section \ref{sec:system-design} and Section \ref{sec:economic-and-governance-model}
hold even under partial compromise of network components.

\section{Conclusion}\label{sec:conclusion}

HOT Protocol enables smart contracts to securely control
their own cryptographic keys
using threshold Multi-Party Computation (MPC) within
Trusted Execution Environments (TEE)
and coordinated on-chain governance.
It allows contracts to authorize and produce signatures
for external actions, forming the basis for Chain-Abstracted
Applications operating assets across multiple networks.

Example use cases include:
\begin{itemize}
	\item Bitcoin-style multi-signature wallets without new
	      on-chain deployments for each policy

	\item Two-factor (2FA) authorization

	\item Seed phrase rotation (if a seed phrase is used to authorize a specific MPC key, this authorization seed can itself be rotated)

	\item Wallet recovery flows

	\item Passkey-based wallets

	\item Other high-assurance authorization policies that are costly
	      to implement with traditional smart wallets

	\item Chain-abstracted dApps capable of managing assets across multiple chains

\end{itemize}


\bibliographystyle{unsrtnat}
\bibliography{main}

\appendix

\section{Passkey Bitcoin Wallet}\label{sec:passkey-bitcoin-wallet}

A passkey is a passwordless authentication method based on public-key
cryptography, where a private key stays securely on your device and
a public key is registered with the service.
It lets you sign in by verifying a cryptographic signature instead of
entering a password, making logins both simpler and more secure.

To demonstrate the use of HOT Protocol,
we present an implemented application that enables passkey-based
Bitcoin wallets — that is, wallets where transaction signing is controlled
solely by the user’s device passkey, itself secured
by fingerprint or facial recognition.

Deployed on NEAR Protocol, \texttt{passkey.auth.hot.tg} maps
each \texttt{key\_id} (Bitcoin wallet) to its passkey \texttt{public\_key}.
It serves as the key-owner contract with the following
\texttt{hot\_verify} method:

\begin{figure}[H]
	\begin{lstlisting}[language=C, caption={Authorization function signature},label={lst:lstlisting2}, escapeinside={(*@}{@*)}]
fn hot_verify(
	message: String,
	key_id: bytes32,
	_metadata: bytes,
) -> bool {
	ecdsa_secp256r1.verify(message, passkey[key_id])
}
				\end{lstlisting}\label{fig:figure3}
\end{figure}

The initialization flow for a user is as follows:
\begin{enumerate}
	\item Obtain the passkey \texttt{public\_key} from their device;

	\item Reserve a \texttt{key\_id} via the MPC Network;

	\item Bind the \texttt{key\_id} to the
	      \texttt{public\_key} in the registry contract.
\end{enumerate}

\end{document}